*Report*

# Unexpectedly large delocalization of the initial excitation in photosynthetic light harvesting.

Khadga J. Karki[1*], Junsheng Chen[1*], Atsunori Sakurai[2*], Qi Shi[1], Alastair T. Gardiner[3], Oliver Kühn[4], Richard J. Cogdell[3] & Tönu Pullerits[1]

[1] Chemical Physics and NanoLund, Lund University, Box 124, 22100 Lund, Sweden

[2] Institute of Industrial Science, The University of Tokyo, 4-6-1 Komaba, Meguro, Tokyo 153-8505, Japan

[3] Division of Biochemistry and Molecular Biology, Institute of Biomedical and Life Sciences, University of Glasgow, Glasgow G12 8QQ, United Kingdom

[4] Institute of Physics, University of Rostock, Albert-Einstein-Str. 23-24, 18059 Rostock, Germany

**Electronic 2D spectroscopy allows nontrivial quantum effects in chemistry and biology to be explored in unprecedented detail. Here, we apply recently developed fluorescence detected coherent 2D spectroscopy to study the light harvesting antenna 2 (LH2) of photosynthetic purple bacteria. The method utilizes the destructive interference between two signal components thereby uncovering cross peaks which are not visible in conventional photon-echo based 2D and transient absorption measurements. Analyses of signal generating quantum pathways leads to the conclusion that, contrary to the currently prevailing physical picture, the two weakly-coupled pigment rings of LH2 share the initial electronic excitation leading to quantum mechanical correlation between the two clearly separate bands. These results are general and have consequences for the interpretation of excited states not only in photosynthesis but in all light absorbing systems. The initial delocalization could be the key for enhancing the light harvesting efficiency via biased motion towards the energy funnel.**

The primary processes in photosynthesis run with nearly 100 % quantum efficiency – almost every absorbed photon leads to a charge separation event. How such high efficiency is achieved and the possible role of quantum processes in it, is currently at the center of active scientific research (*1–4*). One of the possible functional elements of such quantum behavior and optimization in photosynthesis is delocalization – the spatial domain coherently covered by the excited state after light absorption (*5*).

The peripheral light harvesting antenna 2 (LH2) of photosynthetic purple bacteria consists of two rings of bacteriochlorophyll (BChl) *a* molecules embedded into a protein scaffold forming a barrel-like structure (see Fig. 1) (*6*). The rings are responsible for two absorption bands which in most of the LH2s are at 800 nm and 850 nm. Correspondingly, both the rings and the absorption bands are called B800 and B850. In the LH2 of *Rhodopseudomonas (Rps.) acidophila* the B800 ring has 9 and the B850 ring 18 BChl *a* molecules. While the B800 BChls are well separated from each other and from the BChls of the B850 ring, the later form a closely packed aggregate where inter-pigment excitonic interactions are significant (300-400 $cm^{-1}$).

---

[*] These authors contributed equally to this work

In coherent two-dimensional (2D) spectroscopy the spectral information is spread over multiple dimensions revealing features that otherwise are hidden behind broad one-dimensional spectra. Furthermore, since spectral resolution of the method is obtained by temporal scanning of the laser pulses and taking the Fourier transform over the delay time, simultaneous high energy and time resolution beyond the Fourier limit, can be achieved. Therefore, the technique enables ultrafast processes triggered by absorption of light to be followed in detail that is not available by other more conventional time-resolved methods. Various flavors of 2D spectroscopy have emerged. In the original implementation, the third order polarization generated by three laser pulses emits a coherent photon echo-type signal (PE2D) which is mixed with a fourth pulse for phase sensitive heterodyne detection (*7*). The signal depends on the timing of the pulses and is measured in a certain phase-matched direction, which is related to the k-vectors of the incoming laser beams. In the third order nonlinear response one can distinguish rephasing, nonrephasing and double quantum coherence (DQC) signals.

In a more recent development, various incoherent action signals have been applied to measure coherent spectra. Photocurrent detected 2D spectroscopy has provided valuable information about photoinduced processes in quantum well and quantum dot based materials (*8*, *9*). Fluorescence detected 2D (FD2D) spectroscopy was used to investigate the conformation of molecular dimer complexes (*10*). In all these approaches four collinear laser pulses bring the system to an excited state, which can generate the signals as photocurrent or fluorescence. Such incoherent signals do not carry directionality of the phase matching. Instead phase cycling or phase modulation of the four pulses is used to separate the different signal contributions.

Here, we apply FD2D spectroscopy with phase modulation and use the generalized lock-in method to disentangle the different signal components due to the light absorbing states in LH2 complexes from *Rps. acidophila* (see Fig. 1).

Earlier, PE2D spectroscopy has been applied to study the excited states and energy transfer in LH2 (*11*, *12*). A generic 2D spectrum of a system like LH2, has two diagonal peaks corresponding to the two linear absorption bands and possibly also two cross peaks reflecting the correlations between the two bands (see Fig. 2). The PE2D lineshape of the B800 diagonal peak at 77 K suggests that the ring has a significant excitonic character (*13*) despite of the weak electronic couplings between the BChl *a* molecules giving support to earlier theoretical suggestions (*14*). In another study oscillations of the 2D signal as a function of the population time between the excitation pulses and the detection pulse were studied (*15*). The 90 degrees shift between oscillations recorded in different areas of the 2D spectra was taken as evidence for coherent exciton dynamics between B800 and B850.

Indeed, coherent dynamics would lead to such phase shift. However, the phase shifts in 2D beatings can have many other origins (*16*, *17*). An important feature of all previous PE2D results is the lack of any appreciable B800-B850 cross peak at very early population time. This agrees with multi-color pump probe spectroscopy results where the B850 bleach signal grows in from low levels after B800 excitation and the dynamics has been interpreted as excitation transfer from one ring to the other (*18*).

In contrast to these earlier results, here we report FD2D spectra showing prominent cross-peaks at zero population time (Fig 3). In general, cross-peaks in 2D spectroscopy report correlated dynamics and coupling between the involved states. Why does the signature of correlations appear here and not in the previous experiments? In order to answer this question, we need to take a closer look at the different Liouville space pathways contributing to the third order response function of 2D experiments.

The different pathways can be classified as ground state bleach (GSB), stimulated emission (SE) and excited state absorption (ESA). There is a one-to-one correspondence between all possible GSB and SE pathways in PE2D and the pathways in FD2D (*10*). However, for each ESA pathway in the conventional 2D there are two possible ESA pathways in FD2D. One of the pathways ends at a singly excited level $|\alpha\rangle$ or $|\beta\rangle$ while the other at the doubly excited level $|\sigma\rangle$. According to the Feynman diagram rules (*19*) the signals from these two pathways have opposite signs while the strength of them only depends on the yield of fluorescence from the doubly excited state $|\sigma\rangle$ relative to the singly excited state. The doubly excited state in antenna complexes is known to rapidly relax to a singly excited state via nonradiative exciton-exciton annihilation. In LH2 this process is faster than 1 ps (*20*) and should not be confused with the diffusion-driven slow annihilation in large molecular systems. Since the singly excited state lifetime is about 1 ns, the difference of the amount of fluorescence while starting from $|\sigma\rangle\langle\sigma|$ and from $|\alpha\rangle\langle\alpha|$ is negligible. This means that the signals from these two pathways cancel out each other. Therefore, in FD2D of light harvesting complexes and in any other excitonic system with efficient annihilation the ESA signal vanishes. On the other hand, in PE2D spectra of weakly coupled systems with significant energy gap as B800-B850, the ESA signal coincides with the GSB at the cross peaks while having opposite signs, therefore they largely cancel each other (see SI). We propose that this cancellation is the main reason why in PE2D measurements no significant B800-B850 cross peak has been visible at zero population time. This also means that the pump-probe signal of the B850 band after B800 excitation has to be seen as decay of the ESA signal due to the B800 to B850 transfer while the bleach was created at the instance of excitation but was almost entirely canceled by the opposite ESA.

An obvious conclusion from these findings is that the B800 and B850 transitions are correlated. This is to say that, at the moment of excitation, the LH2 acts as a supercomplex, even though the subsequent B800 to B850 transfer is considered to be of incoherent generalized Förster type rate process. The cross peak gives a glimpse to a transient regime, which is usually neglected in the rate model. Interestingly, recent theoretical studies of the coherent properties and entanglement of various excitation scenarios in LH2 also found that the B800 excitonic excitation has delocalization that extends to B850 ring (*21*). Significant delocalization-related transition strength redistribution from antenna to the RC has been predicted despite of very weak interaction between the two (*22*).

Our conclusions are based on the cross peaks which are visible at population time zero because of the perfect cancellation of the ESA signal in FD2D of LH2. A rigorous discussion of this effect based on nonlinear response functions can be found in Ref. (*23*). Recently, action-detected 2D signals from independent particles has been discussed in literature (*24*, *25*). In the present case, could the cross peaks emerge from independent B850 and B800 transitions? The action signal is recorded over a much longer timescale than the four-pulse sequence which is exciting the sample. If during that time the second order populations generated by pulse pairs at different molecules undergo a nonlinear process such as a bimolecular reaction, so-called incoherent mixing can take place providing a signal which carries the same modulation components as the fourth-order signal (*24*). In the context of light harvesting antenna complexes diffusion-driven excitation annihilation can play a role of such a mixing. In SI we show that the incoherent mixing would lead to the signal with opposite sign compared to the GSB signal in case of ESA cancellation scenario. This has consequences to how does the lower cross peak evolve while the SE grows in due to the B800 to B850 transfer during the population time $t_2$. In Fig 3 we compare the corresponding experimental points with the expected dependence of the two scenarios assuming that the B800 to B850 transer rate is 700 fs (*18*). Clearly, the ESA cancellation scenario is supported by the experiment. It has been theoretically predicted that even without incoherent mixing, two molecules can provide fluorescence with the modulation combination frequencies where both excited molecules contribute (*25*). By comparing the signals at $\phi_{21} = \phi_2 - \phi_1$ and $2\phi_{21}$ we demonstrate that our experimental results cannot be explained by such mechanism (for the details see SI).

An independent evidence for our conclusions can be obtained from the DQC signal (*26*). DQC using FD2D was proposed previously (*27*), and was recently measured in a dye molecule (*28*). In our experiments, we have phase locked all the beams, which allows us to phase synchronously detect DQC signal simultaneously with the rephasing and nonrephasing signals (cf. Fig.1). We use the conventional 2D Fourier scanning protocol (*29*) and not the double-frequency scanning as in earlier PE2D DQC implementations (*30*). We observe distinct cross peaks, see Fig 4. In the Liouville pathway analyses we can pair up most of the diagrams so that they take out each other (see SI). The two remaining nondiagonal DQC components each originates from a single Liouville pathway which is in a $|\alpha\beta\rangle\langle 0|$ double quantum coherence during the $t_2$ evolution while it evolves during $t_1$ and $t_3$ in $|\alpha\rangle\langle 0|$ and $|\beta\rangle\langle 0|$ coherences. This signal confirms that the B800 and B850 act collectively, otherwise the B800 and B850 coherences would not be correlated.

Many conceptual discussions have suggested that exciton delocalization may be an important factor in optimizing energy transfer over large distances (*1*, *31*). From the point of view of photosynthesis, the energy transfer within a single LH2 complex is clearly of secondary importance compared to the transfer between the LH2s and from LH2 to the core antenna linked further to the reaction centre (RC). The coupling between the B800 and B850 electronic transitions is about 20-30 $cm^{-1}$ (*32*). We have concluded that B800 and B850 transitions are correlated despite of the rather weak coupling between them. The coupling between LH2s and between LH2 and the core antenna is estimated to be still about 5 times weaker (*33*). Is this coupling enough to create inter-complex delocalization and where is the boarder-line from where the complexes can be considered electronically independent? Answers to these questions may call for rethinking of the ways how the primary energy transfer in photosynthesis works and the current study is the first step in this direction. It has been argued that large delocalization can allow sensing of the energy funnel towards the RC providing biased motion in right direction thereby making energy harvesting more efficient (*31*). Such biased random walk can reach significantly larger distances than a Brownian walk (*34*). In this context the coherent supercomplex character of light harvesting systems may be the key for efficient energy transfer towards the RC.

**Data availability** The datasets generated during and/or analysed during the current study are available from the corresponding author on reasonable request.

**Supplementary information** is available in the online version of the paper.

**Acknowledgements** This work was financed by the Swedish Research Council (VR), the Knut and Alice Wallenberg Foundation, the Swedish Energy Agency, NanoLund, the Crafoord Foundation, Laserlab-Europe EU-H2020 654148 grant. QS acknowledges scholarship support from the China Scholarship Council. ATG and RJC gratefully acknowledge funding from the Photosynthetic Antenna Research Center, an Energy Frontier Research Center funded by the DOE, Office of Science, Office of Basic Energy Sciences under Award Number DE-SC 0001035. AS acknowledges the the Scandinavia-Japan Sasakawa Foundation (No. 14-22) and the Grant-in-Aid for JSPS Research Fellow (16J04694). TP and OK acknowledge financial support from NPRP grant # NPRP7-227-1-034 from the Qatar National Research Fund (a member of Qatar Foundation).

**Author Contributions** T.P. coordinated the project, K.J.K. implemented the experimental setup, A.S., J.S. and K.J.K. performed the experiments and data analyses. A.T.G. and R.J.C. prepared the light harvesting complexes, Q.S. and T.P. performed the Feynman diagram analyses, T.P., K.J.K. and O.K. formulated the conceptual conclusions. All authors contributed to discussion of the data and writing of the manuscript.

**Author Information** The authors declare no competing financial interests. Readers are welcome to comment on the online version of the paper. Correspondence and requests for materials should be addressed to T.P. (tonu.pullerits@chemphys.lu.se).

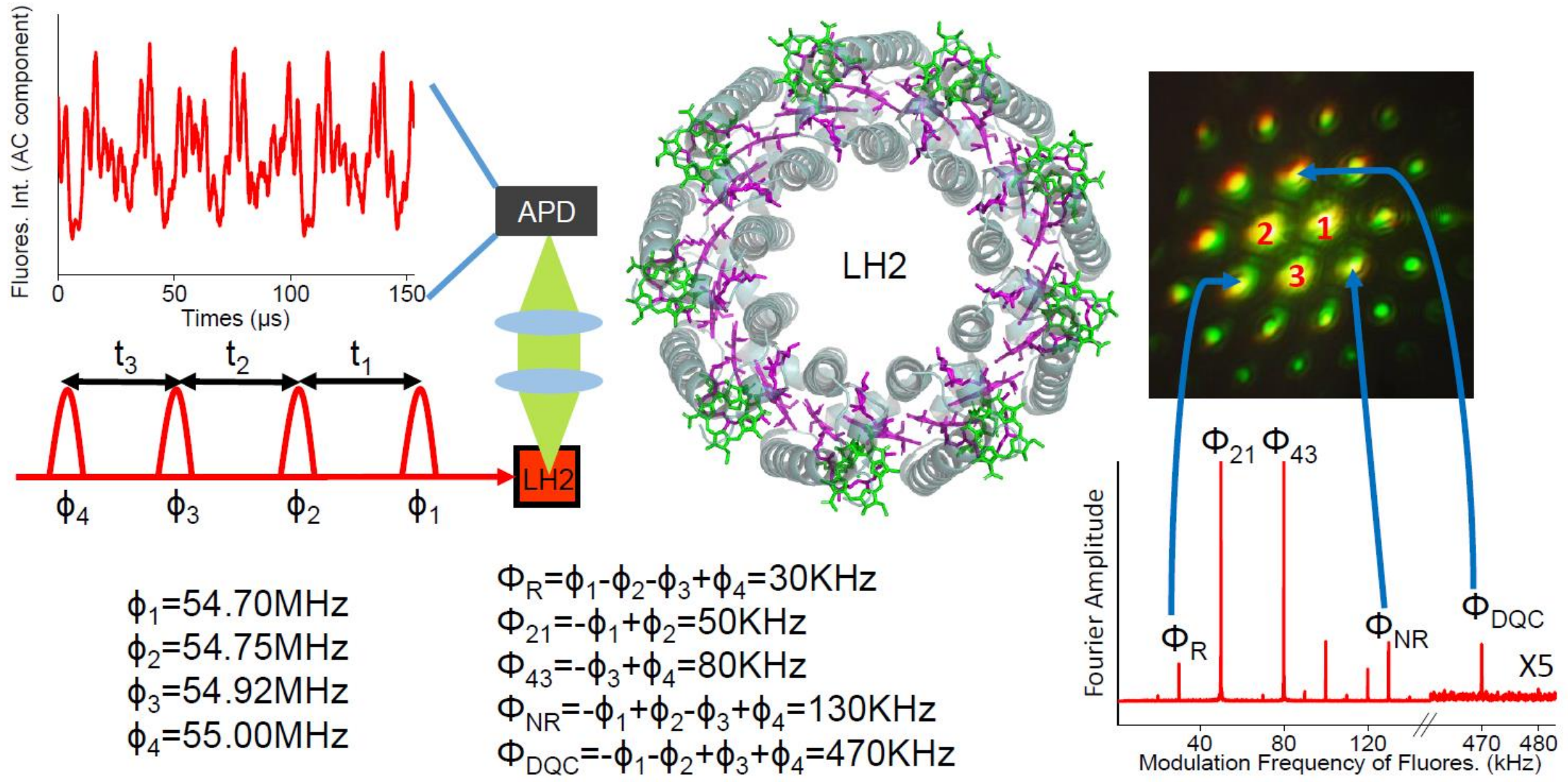

**Figure 1**. Schematic diagram of the FD2D experiment. Four collinear 8-10 fs pulses have tunable delays and different phase modulation frequencies $\phi_i$ . The avalanche photodiode (APD) detects fluorescence from the LH2 sample. The structure of the LH2 is also shown with BChl *a* from different rings shown in green and purple. An example of the oscillatory component of the signal is provided at the upper left corner. In the lower right the Fourier transformed signal is shown. The rephasing, the nonrephasing and the DQC signals appear at $\Phi_R$ = 30 kHz, $\Phi_{NR}$ = 130 kHz and $\Phi_{DCQ}$ = 470 kHz, respectively. The other peaks correspond to other possible pulse combinations which excite fluorescence. For example, 80 kHz is excited by pulses 3 and 4. Also shown is the image of the phase-matched signals in conventional 3-pulse photon echo spectroscopy. 1, 2 and 3 are the three laser beams which have passed the sample. All other spots are phase-matched photon echo signals. The correspondence between the three signals used in the current work is indicated by arrows.

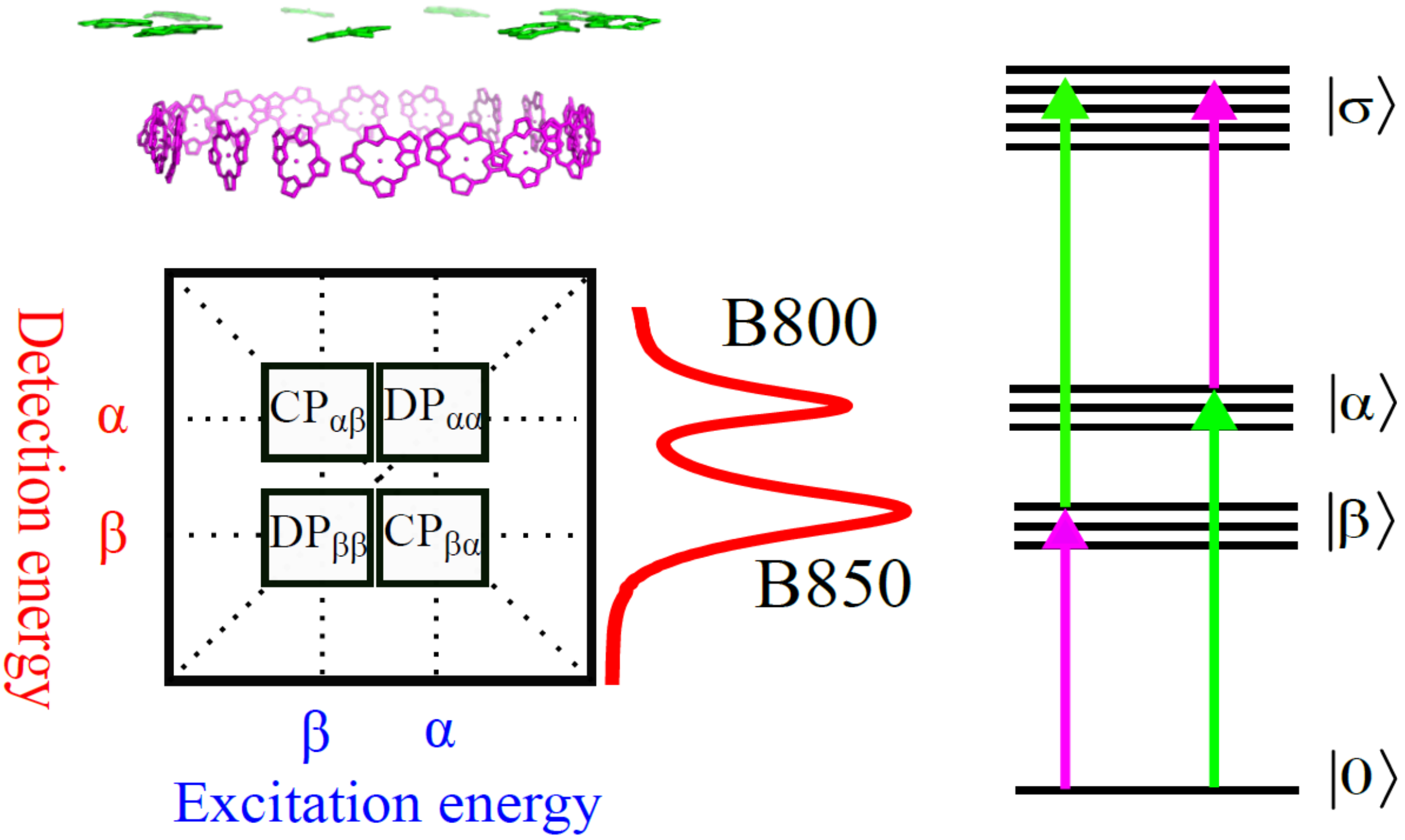

**Figure 2**. Upper left. Arrangement of BChl *a* pigments in the LH2 structure. In purple and green molecules are commonly related to the B850 and B800 bands, respectively. Lower left. Generic 2D spectrum with two diagonal peaks $DP_{\beta\beta}$ and $DP_{\alpha\alpha}$ corresponding to the B850 and B800 bands, respectively (see also the absorption spectrum of the LH2 to the right from the 2D panel). The cross peaks $CP_{\alpha\beta}$ and $CP_{\beta\alpha}$ are shown too. At right hand side the electronic excited state level schema of LH2 to describe the third order optical experiments like 2D spectroscopy, is shown. The one-exciton bands α and β correspond to B800 and B850 bands while the two-exciton band σ correspond to all possible combination of double excitations (including molecular double excitation) in the system. The purple and green arrows correspond to the transitions which mainly involve the β and α transitions, respectively.

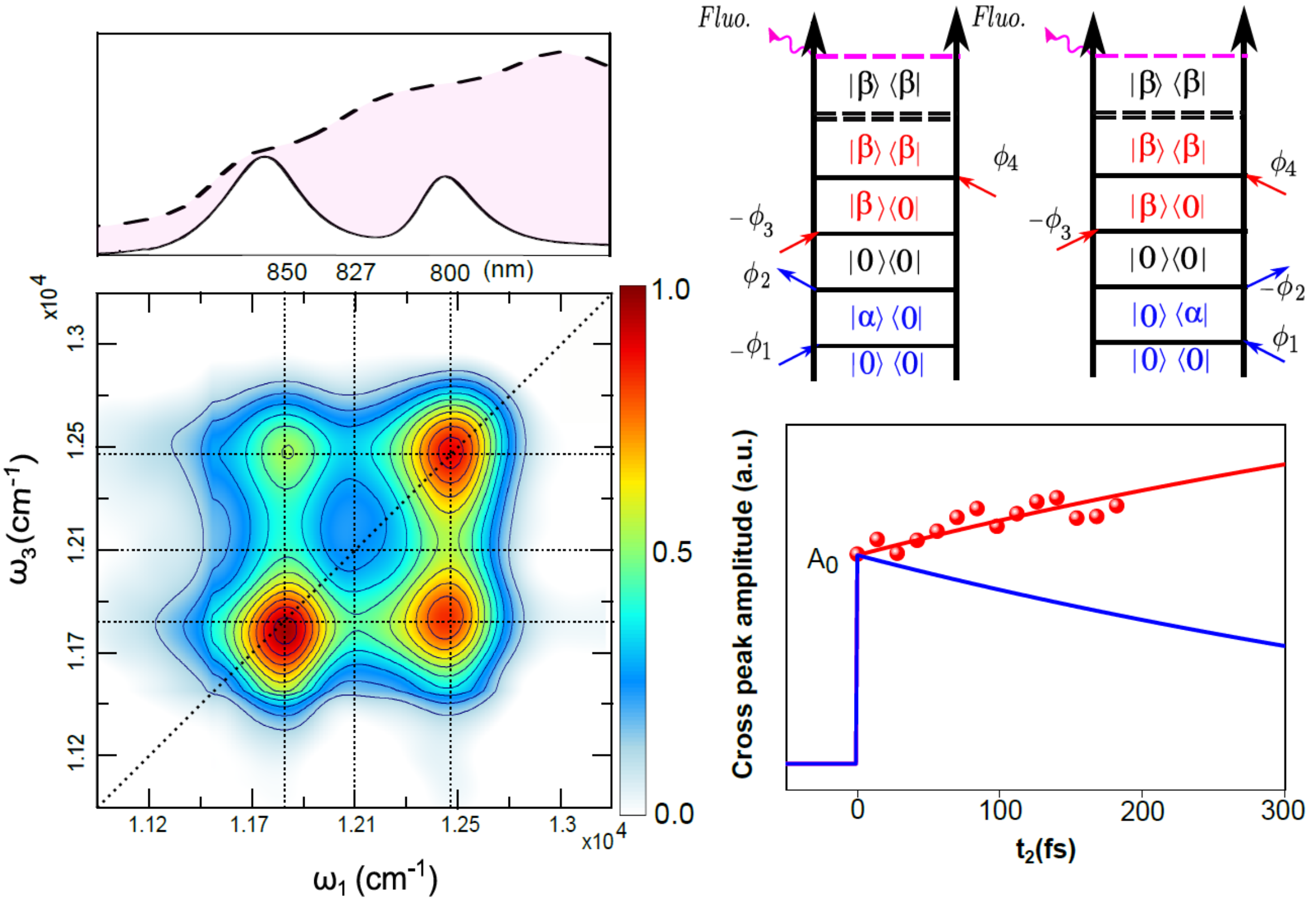

**Figure 3.** Left: total (rephasing + nonrephasing) FD2D of LH2, at $t_2$ = 0 fs. Upper panel shows the absorption spectrum and the spectrum of the pulses (dashed) used in the experiment. In the upper right non-rephasing and rephasing GSB Feynman diagrams of the lower cross peak are shown. Lower right shows the experimental (red points) $t_2$ dependence of the lower cross peak amplitude together with the expected dependence due to the B800 to B850 transfer based on the ESA cancellation argument, red line, and the independent bimolecular annihilation (incoherent mixing) argument, blue line.

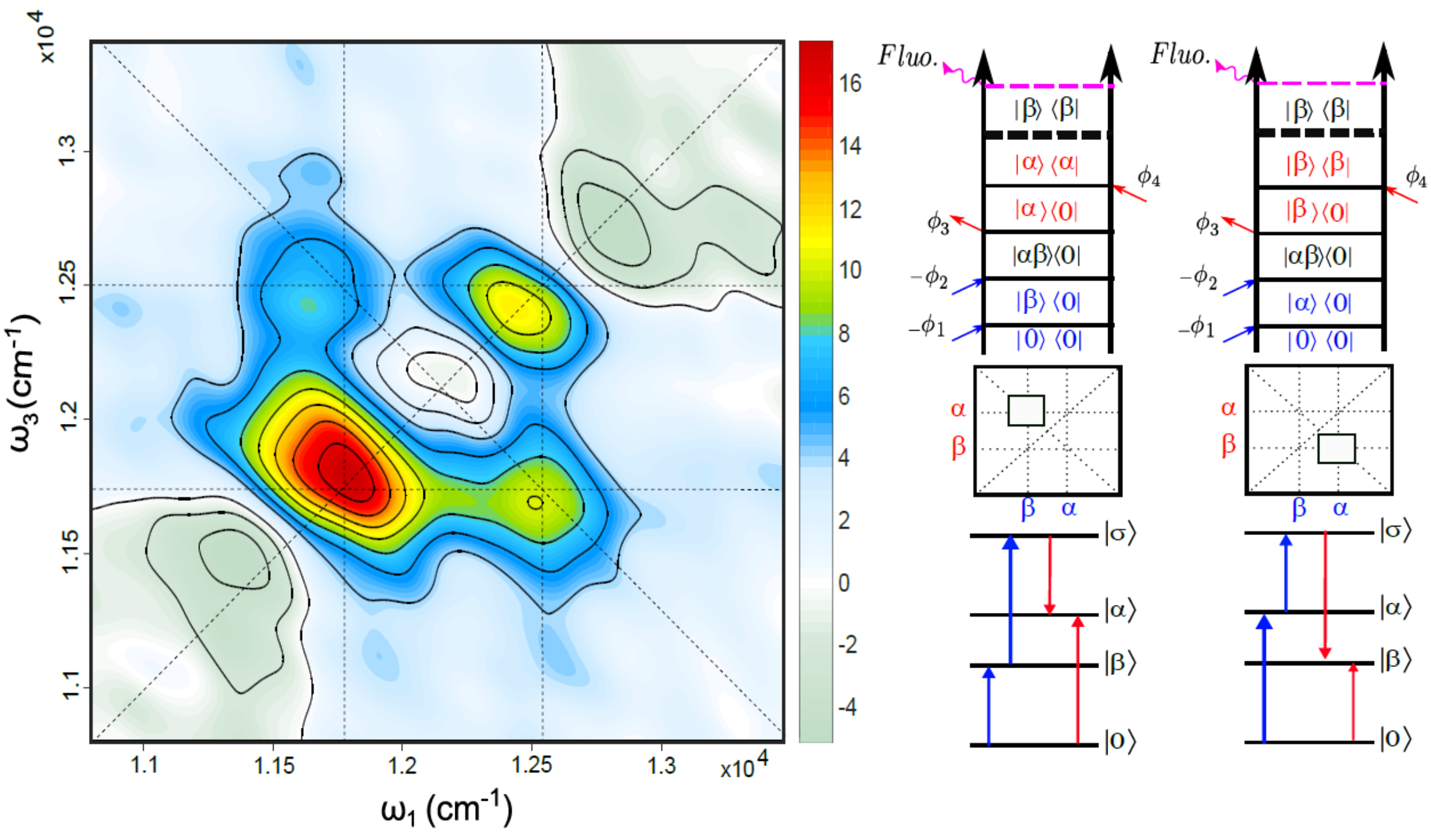

**Figure 4**. Left: Double quantum coherence 2D spectrum. Right: two pathways which lead to the nondiagonal bands (cross-peaks). Here the doubly excited state corresponds to the configuration where both B800 and B850 are excited: $|\sigma\rangle = |\alpha\beta\rangle$ .